\documentclass{PoS}

\title{Threshold and jet radius joint resummation for single-inclusive jet production}

\ShortTitle{Threshold and jet radius joint resummation}

\author{\speaker{Sven-Olaf Moch}\\
  II. Institut f\"ur Theoretische Physik, 
  Universit\"at Hamburg, 
  Luruper Chaussee 149, 
  D-22761 Hamburg, Germany\\
  E-mail: \email{sven-olaf.moch@desy.de}}

\author{Engin Eren \\
  Deutsches Elektronen Synchrotron DESY,
  Notkestra{\ss}e 85,
  D-22607 Hamburg, Germany\\
  E-mail: \email{engin.eren@desy.de}}

\author{Katerina Lipka \\
  Deutsches Elektronen Synchrotron DESY,
  Notkestra{\ss}e 85,
  D-22607 Hamburg, Germany\\
  E-mail: \email{katerina.lipka@desy.de}}

\author{Xiaohui Liu \\
  Center of Advanced Quantum Studies, 
  Department of Physics, 
  Beijing Normal University, 
  Beijing 100875, China\\
  E-mail: \email{xiliu@bnu.edu.cn}}

\author{Felix Ringer\\
  Nuclear Science Division, 
  Lawrence Berkeley National Laboratory, 
  Berkeley, California 94720, 
  USA\\
  E-mail: \email{fmringer@lbl.gov}}

\abstract{The QCD predictions for single-inclusive jet production are 
computed with joint resummation of threshold and jet radius logarithms. 
The results are compared to those based on fixed order perturbation theory up to next-to-next-to-leading order
and to data by the CMS collaboration measured in proton-proton collisions at the LHC at $ \sqrt{S}=8$~TeV.
The joint resummation results are in remarkable agreement with the data.}

\FullConference{Loops and Legs in Quantum Field Theory (LL2018)\\
		29 April 2018 - 04 May 2018\\
		St. Goar, Germany}

\begin{document}

The hadro-production of jets in proton-proton collisions at the Large Hadron
Collider (LHC) is one of the basic processes which exposes all features 
of the gauge theory of the strong interactions, quantum chromodynamics (QCD).
Within standard QCD factorization the description of jet hadro-production 
starts off at Born level as a hard $2 \to 2$ scattering reaction of partons 
in the incident protons and its theory prediction is directly proportional to
powers of the strong coupling constant $\alpha_s$ and to the parton luminosity,
i.e., the distribution of the fractions of parton momenta relative to those of
the colliding protons. 
These characteristics make jet hadro-production a very interesting process for
determinations of both, the value of the strong coupling $\alpha_s$ as well as 
the parton distribution functions (PDFs) from collider data, 
see for instance Refs.~\cite{Accardi:2016ndt,Britzger:2017maj}.

The single-inclusive jet production cross section, 
i.e., the observable of interest for us here, 
is obtained by summing over all jets that are observed in an event. 
The corresponding double differential expression for a jet of cone size $R$ reads 
\begin{eqnarray}
  \label{eq:crs}
  \frac{p_T^2 \mathrm{d}^2 \sigma    }{   \mathrm{d}p_T^2 \mathrm{d} y } 
  & = & \sum_{i_1i_2} 
  \int_0^{V(1-W)}  \!\! \mathrm{d} z 
  \int_{\frac{VW}{1-z}}^{1- \frac{1-V}{1-z}} \!\! \mathrm{d}v \,
  x_1^2\,  f_{i_1}(x_1) \, x_2^2 \,  f_{i_2}(x_2)\, 
  \frac{\mathrm{d}^2 \hat{\sigma}_{i_1i_2}}{ \mathrm{d} v \, \mathrm{d} z } (v,z,p_T,R)   \,,  
\end{eqnarray}
where $p_T$ and $y$ denote the transverse momentum and rapidity of the jet.
The sums $i$ run over all partonic channels which contribute through 
the convolution of the hard partonic cross sections $\hat{\sigma}_{i_1i_2}$
with PDFs $f_{i}$ evaluated at the fractions $x_1 = VW/v/(1-z)$ and $x_2 = (1-V)/(1-v)/(1-z)$ 
of the proton momenta, where $V = 1- p_T e^{-y}/\sqrt{S}$, $VW = p_T e^{y}/\sqrt{S} $ and $\sqrt{S}$ is the hadronic center-of-mass energy. 
The partonic kinematic variables are $s = x_1 x_2 S$, $v = u/(u+t)$ and $z=s_4/s$,
defined in terms of the standard Mandelstam variables $t$, $u$ and, respectively,
the invariant mass $s_4$ of the partonic system recoiling against the observed jet, cf.~Ref.~\cite{Liu:2017pbb}.

The respective higher order perturbative QCD predictions to the 
partonic cross sections $\hat{\sigma}_{i_1i_2}$ in Eq.~(\ref{eq:crs}) are known to
next-to-leading order (NLO) since long~\cite{Ellis:1992en,Nagy:2001fj} 
while those at next-to-next-to-leading order (NNLO) have been computed
recently in the leading-color approximation~\cite{Currie:2016bfm}, 
that is for large values of $N_c$ for a general SU$(N_c)$ gauge group with 
additional corrections parametrically suppressed as $1/N_c^2$.
This is supposed to approximate the full NNLO calculation very well. 

Higher order corrections to hard processes in QCD are generally expected 
to display an apparent convergence of the perturbative expansion as well as 
to show a significant reduction of the dependence on the scales $\mu_R$ 
for ultraviolet renormalization and $\mu_F$ for QCD factorization. 
Both these features lead to a stabilization of the theoretical predictions.
However, comparisons of the fixed order NNLO results of Ref.~\cite{Currie:2016bfm}
for the single-inclusive jet production cross section with some of the LHC data have not
been entirely satisfactory~\cite{Currie:2017ctp,ATLAS:2017ifd}.
Moreover, the NNLO corrections do change significantly depending on the chosen values for the hard scales $\mu_R$ and $\mu_F$ 
(denoted here collectively by $\mu$).
Choices like $\mu=p_T^{\rm{max}}$ with the natural hard scale 
of the transverse momentum $p_T^{\rm{max}}$ of the leading jet or, alternatively, 
scales like $\mu=p_T$ with the transverse momentum $p_T$ of each individual jet in the event
lead to completely different theoretical predictions. 
The latter choice $\mu=p_T$ typically involves much softer scales because 
kinematical configurations in events with three or more hard jets 
or events with hard emissions outside the jet fiducial cuts such as smaller
jet cone radii $R$ generate a hierarchy in the transverse momenta 
between the leading and subleading jets in the event, $p_T^{\rm{max}} = p_{T,1} \gg p_{T,2} \geq p_{T,3} \dots$, 
cf. Ref.~\cite{Currie:2018xkj}.

This infrared sensitivity implies the existence of 
large higher order corrections beyond fixed order in perturbation theory and
merits a short discussion of their origins. 
Large logarithms in QCD perturbation theory arise systematically 
from the cancellations of infrared divergences between real and virtual corrections
near some boundary of the phase space.
For single-inclusive jet production this is realized near threshold for large
$p_T$, when the event kinematics are almost Born-like 
and double logarithms appear in the partonic cross section at $n$-th order 
as $\alpha_s^n(\ln^k(z)/z)_+$ where $k\le 2n-1$ and $z$ is the measure for the
distance from partonic threshold, cf. Eq.~(\ref{eq:crs}).
On the other hand, the definition of jets via their cone sizes $R$ as an external quantity 
introduces large single-logarithmic corrections in the partonic cross section 
at $n$-th order as $\alpha_s^n\ln^k(R)$ with $k\le n$.

Both, threshold and small-$R$ logarithms require an all-order resummation 
and one expects competing effects from these two sources.
The resummation of threshold logarithms leads to an enhancement of the cross section for large
$p_T$ and has been studied in Refs.~\cite{Kidonakis:1998bk,Kumar:2013hia,deFlorian:2013qia}.
On the contrary, the resummation of small-$R$ logarithms alone leads to a decrease of the 
cross section in the entire range of $p_T$~\cite{Dasgupta:2016bnd,Kang:2016mcy}.
A framework for the joint resummation of both, threshold and jet radius
logarithms simultaneously, has recently been developed in Ref.~\cite{Liu:2017pbb}.
To that end the partonic cross sections $\hat{\sigma}_{i_1i_2}$ in
Eq.~(\ref{eq:crs}) are further factorized within the soft collinear effective theory (SCET)~\cite{Becher:2015hka}
as
\begin{eqnarray}
  \label{eq:fac}
  \frac{\mathrm{d}^2 \hat{\sigma}_{i_1i_2}}{ \mathrm{d} v \, \mathrm{d} z } & = &
  s \int \mathrm{d} s_X \, \mathrm{d}s_c \mathrm{d}s_G \, \delta(z s -s_X -s_G - s_c) 
  {\mathrm Tr} \left[   { \bf H}_{i_1i_2}(v,p_T\,, \mu_h\,, \mu) \,  {\bf S}_G(s_G\,,\mu_{sG}\,, \mu) \right] 
  \nonumber \\
  && \hspace*{5mm} \times\,
  J _X(s_X\,, \mu_{X} \,, \mu) 
  \sum_m {\mathrm Tr}\left[  J_{m}(p_T R \,,\mu_J\,, \mu) \otimes_{\Omega} S_{c,m}(s_c R\,, \mu_{sc} \,,  \mu)  \right]  \,, 
\end{eqnarray}
where the specific functions capture the dynamics of individual kinematic regions, i.e., 
the function ${\bf H}_{i_1i_2}$ for the underlying hard $2 \to 2$ scattering, 
the global soft function ${\bf S}_G$ for wide angle soft radiation from partons, 
which cannot resolve the jet with a small radius $R$ and 
the soft collinear (`coft') function $S_{c}(s_c R)$ for soft radiation near the jet boundary. 
Likewise, the inclusive jet function $J_X(s_X)$ describes the recoiling
collimated radiation with invariant mass $s_X$,  
while the signal-jet function $J_{m}(p_T R)$ accounts for energetic radiation inside jet.
The sum in Eq.~(\ref{eq:fac}) runs over all collinear splittings, the traces
are to be taken in color space and `$\otimes$' denotes the associated angular integrals.
The functions, ${\bf H}_{i_1i_2}$, ${\bf S}_G$, $J_X(s_X)$, $S_{c}(s_c R)$ and $J_{m}(p_T R)$ 
in Eq.~(\ref{eq:fac}) are known to NLO at least, which allows to perform 
the joint resummation at next-to-leading logarithmic (NLL) accuracy, by
evolving all functions with their renormalization group equations 
from their natural scales $\mu_i$ to a common hard scale 
$\mu = \mu_h = p_T^{\rm max}$, see Ref.~\cite{Liu:2017pbb} for details.

The SCET factorization used in Eq.~(\ref{eq:fac}) holds in the threshold
regime for $z \to 0$ and for small jet radii, $R \ll 1$ and is 
matched to the fixed order QCD result as follows,
\begin{eqnarray}
  \label{eq:sigma-NLLres}
  \sigma_{\rm NLO + NLL} = \sigma_{\rm NLO} - \sigma_{\rm NLO_{sing}} + \sigma_{\rm NLL} \,,  
\end{eqnarray}
where $\sigma_{\rm NLO}$ ($\sigma_{\rm NLL}$) denotes the NLO fixed order (NLL resummed) result and $\sigma_{\rm NLO_{sing}}$ subtracts the
logarithmically enhanced contributions at NLO in QCD to avoid double counting.

The first phenomenological studies of the joint jet radius and threshold
resummation results with Eq.~(\ref{eq:sigma-NLLres}) and comparisons to LHC data  
have been carried out in Refs.~\cite{Liu:2017pbb,Liu:2018ktv} and have 
shown that the large threshold and the small-$R$ logarithms are indeed
responsible for the bulk of the radiative corrections 
in the kinematic range from moderate to large jet-$p_T$.
Moreover, the effect of small-$R$ resummation is dominant, even for larger
cone sizes, although less pronounced, while threshold logarithms are relevant
for very large values of jet-$p_T$ and large cone sizes.
In particular, the use of the resummed prediction $\sigma_{\rm NLO + NLL}$ 
indicates a clear systematic improvement in the description at the available data.
It is also worth emphasizing, that in a different framework, 
threshold resummation for jets including the effects of the jet cone size $R$ has recently been studied 
in the parton shower event generator {\it Deductor}~\cite{Nagy:2017dxh}
and the findings agree with ours.

\begin{table}[t!]
\renewcommand{\arraystretch}{1.3}
\begin{center}
\begin{tabular}{|c|c|c|c|}
\hline
\multicolumn{1}{|c|}{data set} &
\multicolumn{1}{c|}{NLO} &
\multicolumn{1}{c|}{NNLO} &
\multicolumn{1}{c|}{NLO $+$ NLL} \\
\hline
$0.0 \le |y| < 0.5$ 
& $36/33$
& $55/33$
& $39/33$
\\
$0.5 \le |y| < 1.0$ 
& $34/32$
& $37/32$
& $35/32$
\\
$1.0 \le |y| < 1.5$ 
& $22/31$
& $32/31$
& $28/31$
\\
$1.5 \le |y| < 2.0$ 
& $10/26$
& $26/26$
& $12/26$
\\
$2.0 \le |y| < 2.5$ 
& $8/19$
& $25/19$
& $9/19$
\\
$2.5 \le |y| < 3.0$ 
& $8/16$
& $15/16$
& $8/16$
\\
\hline
total $\chi^2$/dof
& $142/157$
& $229/157$
& $154/157$
\\
\hline
\end{tabular}
\caption{\label{tab:chi}
The values of $\chi^2$/dof for the QCD theory predictions at NLO, NNLO and 
NLO $+$ NLL accuracy using the CT14 PDF set at NNLO~\cite{Dulat:2015mca}
and the scale choice $\mu_R=\mu_F=p_T^{\rm{max}}$ 
for the single-inclusive jet production cross sections in various rapidity bins 
measured by CMS at $\sqrt{S} = 8$~TeV~\cite{Khachatryan:2016mlc}.
}
\end{center}
\end{table}

Comparisons of jet data collected at the LHC to theory predictions beyond
fixed order NLO in QCD are not yet widespread in the literature. 
This is due to the large computational overhead in the preparation of suitable
fast interfaces for such studies, e.g., {\it fastNLO}~\cite{Britzger:2012bs}.
Here, we continue these phenomenological studies for single-inclusive jet
production data measured by CMS at $\sqrt{S} = 8$~TeV~\cite{Khachatryan:2016mlc}
with the anti-$k_T$ clustering algorithm~\cite{Cacciari:2008gp} for a size parameter of $R=0.7$.
The high statistics CMS data for the double-differential inclusive jet cross section in $p_T$ and $y$ 
correspond to an integrated luminosity of 19.7 inverse femtobarn. 
We confront these data to the QCD theory predictions based on Eq.~(\ref{eq:crs}) 
at NLO, NNLO and NLO $+$ NLL, using Eq.~(\ref{eq:sigma-NLLres}) in the latter case.
The set-up is as follows. The scales are chosen as $\mu_R=\mu_F=p_T^{\rm{max}}$ 
and in all cases, we employ the CT14 PDF set at NNLO~\cite{Dulat:2015mca} with the value of $\alpha_s(M_Z) = 0.1180$,
independent of the perturbative order. 

In Tab.~\ref{tab:chi} we display the values of $\chi^2$/dof (degree of freedom) 
for the respective theory predictions for each rapidity bin. 
The total CMS data set at $\sqrt{S} = 8$~TeV~\cite{Khachatryan:2016mlc} yields 
$\chi^2$/dof=142/157 when compared to the fixed order NLO prediction,  
$\chi^2$/dof=229/157 at NNLO and 
$\chi^2$/dof=154/157 at NLO $+$ NLL.
Thus, the fixed order theory description at NNLO with the hard scale $\mu=p_T^{\rm{max}}$ 
is clearly disfavored, an observation in line with the earlier studies of Refs.~\cite{Currie:2017ctp,ATLAS:2017ifd}.

\begin{figure}[t!]
\begin{center}
\includegraphics[width=0.975\textwidth]{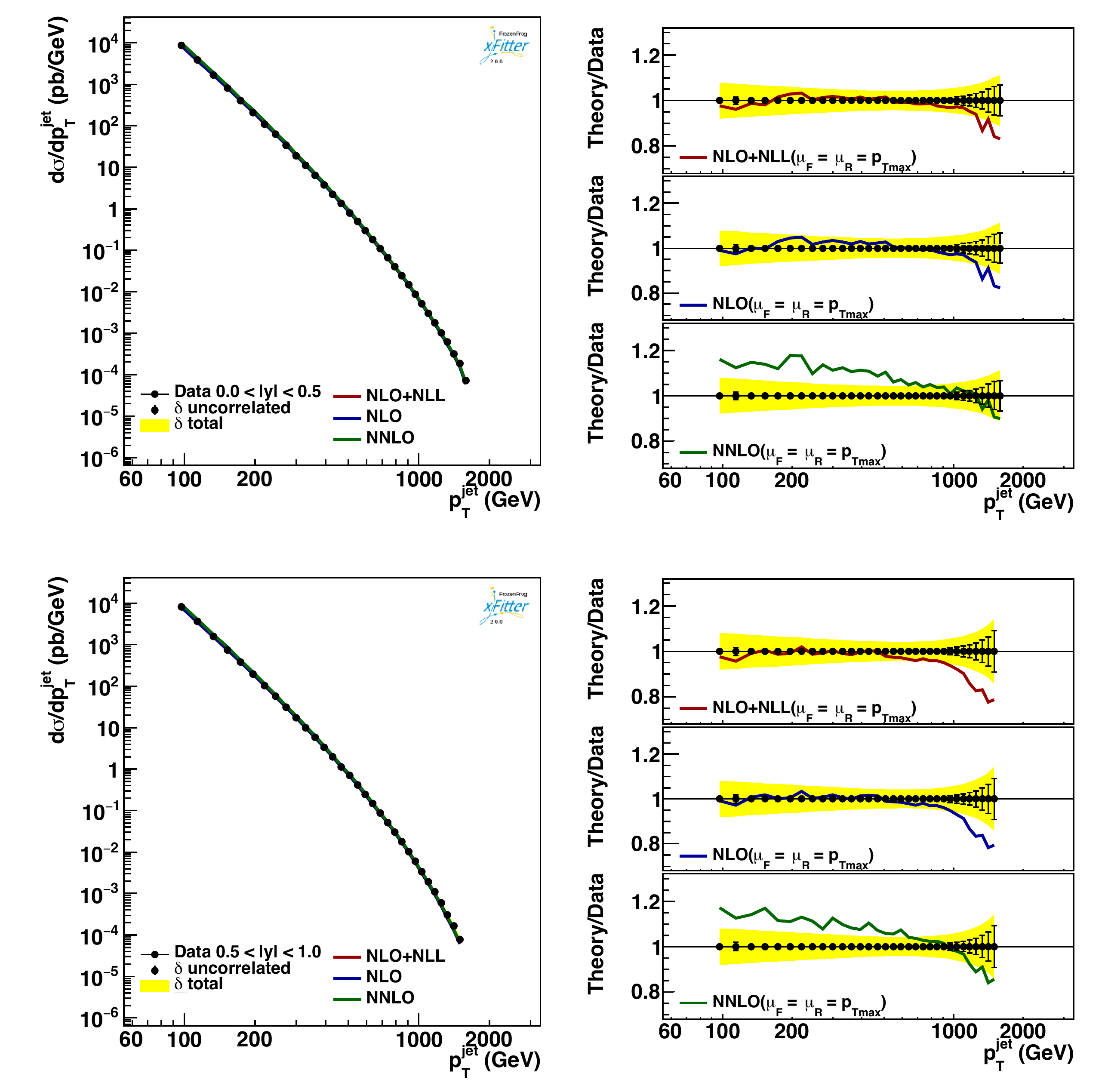}
\caption{\label{fig1} 
The single-inclusive jet production cross section for $pp\to\rm{jet}+X$ as a
function of the jet $p_T$ as measured by CMS at $\sqrt{S} = 8$~TeV~\cite{Khachatryan:2016mlc} 
for a size parameter of $R=0.7$ 
in the rapidity bins $0.0 \le |y| < 0.5$ and $0.5 \le |y| < 1.0$ 
compared to the QCD theory predictions at NLO (blue), NNLO (green) and 
NLO $+$ NLL (red) accuracy using the CT14 PDF set at NNLO~\cite{Dulat:2015mca}
and the scale choice $\mu_R=\mu_F=p_T^{\rm{max}}$.
The left panel shows the absolute cross section,  
the right panel the ratio $\sigma_{\rm Data}/\sigma_{\rm Theory}$.
}   
\end{center}
\end{figure}

\begin{figure}[t!]
\begin{center}
\includegraphics[width=0.975\textwidth]{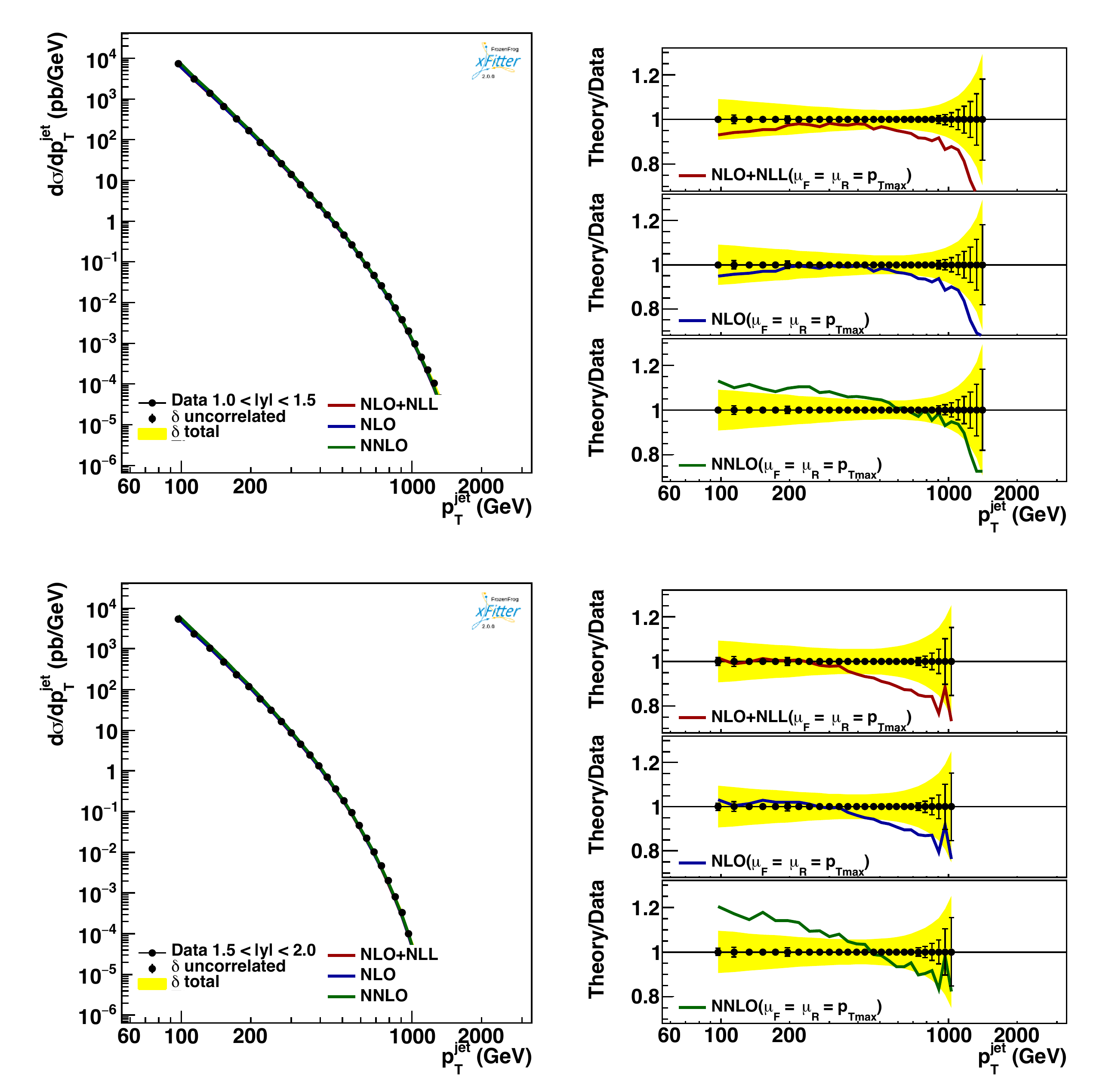}
\caption{\label{fig2} 
  Same as Fig.~\ref{fig1} for the 
  rapidity bins $1.0 \le |y| < 1.5$ and 
  $1.5 \le |y| < 2.0$.
}   
\end{center}
\end{figure} 

\begin{figure}[t!]
\begin{center}
\includegraphics[width=0.975\textwidth]{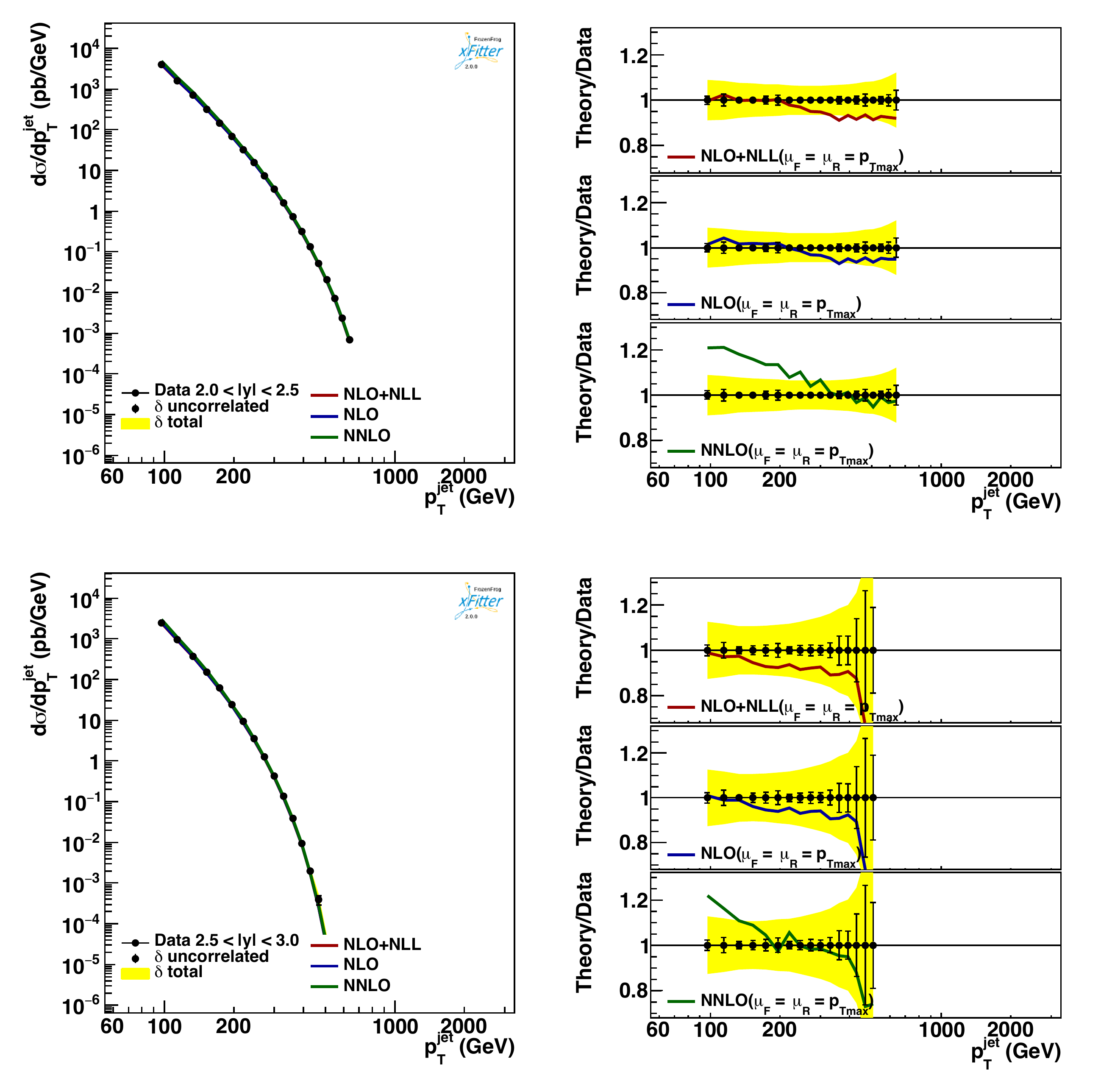}
\caption{\label{fig3} 
  Same as Fig.~\ref{fig1} for the 
  rapidity bins $2.0 \le |y| < 2.5$ and 
  $2.5 \le |y| < 3.0$.
}   
\end{center}
\end{figure}

In Figs.~\ref{fig1},~\ref{fig2} and ~\ref{fig3} we show both, the absolute cross
section for each rapidity bin as well as the separate ratios $\sigma_{\rm Data}/\sigma_{\rm Theory}$
for the theory predictions at NLO, NNLO and NLO $+$ NLL.
While all predictions undershoot the data at the largest values of $p_T$, e.g., 
$p_T \gtrsim 1$~TeV for central rapidities, their individual trends for moderate
values of $p_T$ are completely different and the NNLO results overshoot the
data significantly. 
The joint resummation results on the other hand show remarkable agreement.
It should be mentioned here, that although the jet cone size of $R=0.7$ is already somewhat larger, 
the studies of Refs.~\cite{Liu:2017pbb,Liu:2018ktv} have shown that there is still significant
numerical impact due to the small-$R$ resummation.

While the Figs.~\ref{fig1},~\ref{fig2} and ~\ref{fig3} are based on CT14 PDF set at NNLO~\cite{Dulat:2015mca} 
the findings concerning the improved quality of the joint resummation result
in the description of the data do not significantly depend 
on the PDF choice, as discussed in Refs.~\cite{Liu:2018ktv,Eren:2018}.
In particular, these aspects are investigated in detail in a full QCD
fit~\cite{Eren:2018}, in which both, the PDFs and $\alpha_s$ are determined
simultaneously using single-inclusive jet data by CMS at $\sqrt{S}=8$ and 13~TeV. 
The conclusion is made, that the deficits of the fixed order NNLO predictions to
describe the observed cross sections cannot be mitigated by changes in the PDFs
or the strong coupling.

In summary, we note that the NLO $+$ NLL calculations greatly improve the theoretical predictions. 
However, they do exhibit associated scale uncertainties (not displayed in Figs.~\ref{fig1},~\ref{fig2} and ~\ref{fig3}), 
which are still large~\cite{Liu:2017pbb,Liu:2018ktv} 
and require improving the joint threshold and small-$R$ joint resummation 
to next-to-next-to-leading logarithmic (NNLL) accuracy.
This can be achieved by computing the individual functions 
in SCET factorization in Eq.~(\ref{eq:fac}) to NNLO so that the susequent 
evolution with the renormalization group equations then resums all logarithms to
NNLL. This will be subject of future work.  

\subsection*{Acknowledgments}
The plots have been generated in the {\it xFitter} framework~\cite{Alekhin:2014irh,Bertone:2017tig}.

S.M. acknowledges contract 05H15GUCC1 by BMBF. 
X.L. is supported by the National Natural Science Foundation of China under Grant No. 11775023 
and the Fundamental Research Funds for the Central Universities.
F.R. is supported by the Department of Energy under Contract No. DE-AC0205CH11231, and the LDRD Program of LBNL. 


\providecommand{\href}[2]{#2}\begingroup\raggedright\endgroup

\end{document}